\documentclass[useAMS,usenatbib]{mn2e}
\usepackage{amsmath}\usepackage{graphicx}
\usepackage[papersize={8.5in,11in}]{geometry}
\renewcommand{\vec}[1]{\boldsymbol{\mathrm{#1}}}

\begin{document}

\title{The bending of light and lensing in modified gravity}

\author[J. W. Moffat and V. T. Toth]{J. W. Moffat$^{1,2}$ and V. T. Toth$^1$\\
$^1$Perimeter Institute for Theoretical Physics, Waterloo, Ontario N2L 2Y5, Canada\\
$^2$Department of Physics, University of Waterloo, Waterloo, Ontario N2L 3G1, Canada}

\maketitle

\begin{abstract}
Our modified gravity theory (MOG) was used successfully in the past to explain a range of astronomical and cosmological observations, including galaxy rotation curves, the CMB acoustic peaks, and the galaxy mass power spectrum. MOG was also used successfully to explain the unusual features of the Bullet Cluster 1E0657-558 without exotic dark matter. In the present work, we derive the relativistic equations of motion in the spherically symmetric field of a point source in MOG and, in particular, we derive equations for light bending and lensing. Our results also have broader applications in the case of extended distributions of matter, and they can be used to validate the Bullet Cluster results and provide a possible explanation for the merging clusters in Abell 520.
\end{abstract}



\section{Introduction}

~\par

Our modified gravity theory (MOG), also known as Scalar-Tensor-Vector Gravity (STVG, \cite{Moffat2006a}), is a theory based on an action principle. The action incorporates, in addition to the usual Einstein-Hilbert term associated with the metric $g_{\mu\nu}$, a massive vector field $\phi_\mu$, and three scalar fields representing running values of the gravitational constant $G$, the vector field's mass $\mu$, and its coupling strength $\omega$. The vector field is associated with a fifth force charge that is proportional to mass-energy. This fifth force is repulsive; for a large source mass, at large distances, gravity is stronger than that predicted by Newton or Einstein, but at short range, this stronger gravitational attraction is canceled by the fifth force field, leaving only Newtonian gravity. The theory has been used successfully to account for the rotation curves of galaxies \citep{Moffat2004,Moffat2005,Brownstein2006a,Moffat2007e}, the mass profiles of galaxy clusters \citep{Brownstein2006b}, and cosmological observations \citep{Moffat2007b,Moffat2007c} without exotic dark matter.

The theory was also used to offer an explanation for the unusual features of the Bullet Cluster 1E0657-558 \citep{Brownstein2007}. \cite{Brownstein2007} used a lensing formula that was deduced from the nonrelativistic acceleration law for test particles in the vicinity of a MOG point source.

Deflection of light by gravity is increasingly recognized as an important test for models of dark matter and dark energy as well as modified gravity theories. \cite{Uzan2001} show that a comparison of cosmological lensing data with large scale structure surveys can serve as an effective test of gravity on cosmological scales. \cite{Amendola2008} discuss the possibility that large-scale weak lensing surveys can distinguish between the standard ($\Lambda$CDM) model of cosmology and several modified gravity models, which they characterize by a parameterization of their expansion history and structure growth. Gravitational lensing is also one of the tests proposed by \cite{Jain2008} for modified gravity theories. The specific case of $f(R)$ gravity and weak lensing was studied by \cite{Tsujikawa2008}, while \cite{Thomas2008} study weak lensing in the context of the Dvali Gabadadze Porrati model\footnote{We thank the anonymous referee for providing these references.}.

In the present work, we derive a light bending formula for MOG using a fully relativistic approach, following the route described by \cite{Weinberg1972}. We also develop a formulation for extended distributions of sources, which is important not only for lensing, but also in future, planned work that includes astronomical $N$-body simulations using MOG.

We begin in Section~\ref{sec:theory}, reviewing the basic equations of MOG and the results of an exact numerical solution in the spherically symmetric case. We proceed by developing the equations of motion in the vicinity of a MOG point source in Section~\ref{sec:motion}. In Section~\ref{sec:bending}, we obtain an exact treatment for light bending in the field of a point source. We generalize our discussion to extended sources and lensing in Section~\ref{sec:extended}. Lastly, in Section~\ref{sec:conclusions} we conclude by discussing the possible consequences for the Bullet Cluster and for the merging clusters of galaxies Abell 520 \citep{Mahdavi2007}.

\onecolumn

\section{Modified gravity theory}
\label{sec:theory}

Our modified gravity theory is based on postulating the existence of a massive vector field, $\phi_\mu$. The choice of a massive vector field is motivated by our desire to introduce a {\em repulsive} modification of the law of gravitation at short range. The vector field is coupled universally to matter. The theory, therefore, has three constants: in addition to the gravitational constant $G$, we must also consider the coupling constant $\omega$ that determines the coupling strength between the $\phi_\mu$ field and matter, and a further constant $\mu$ that arises as a result of considering a vector field of non-zero mass, and controls the coupling range. The theory promotes $G$, $\mu$, and $\omega$ to scalar fields, hence they are allowed to run, resulting in the following action \citep{Moffat2006a,Moffat2007e}:
\begin{equation}
S=S_G+S_\phi+S_S+S_M,
\end{equation}
where
\begin{equation}
S_G=-\frac{1}{16\pi}\int\frac{1}{G}\left(R+2\Lambda\right)\sqrt{-g}~d^4x,
\end{equation}
\begin{equation}
S_\phi=-\int\omega\left[\frac{1}{4}B^{\mu\nu}B_{\mu\nu}-\frac{1}{2}\mu^2\phi_\mu\phi^\mu+V_\phi(\phi)\right]\sqrt{-g}~d^4x,
\end{equation}
\begin{equation}
S_S=-\int\frac{1}{G}\left[\frac{1}{2}g^{\mu\nu}\left(\frac{\nabla_\mu G\nabla_\nu G}{G^2}+\frac{\nabla_\mu\mu\nabla_\nu\mu}{\mu^2}-\nabla_\mu\omega\nabla_\nu\omega\right)+\frac{V_G(G)}{G^2}+\frac{V_\mu(\mu)}{\mu^2}+V_\omega(\omega)\right]\sqrt{-g}~d^4x,
\end{equation}
where $S_M$ is the ``matter'' action, $B_{\mu\nu}=\partial_\mu\phi_\nu-\partial_\nu\phi_\mu$, while $V_\phi(\phi)$, $V_G(G)$, $V_\omega(\omega)$, and $V_\mu(\mu)$ denote the self-interaction potentials associated with the vector field and the three scalar fields. The symbol $\nabla_\mu$ is used to denote covariant differentiation with respect to the metric $g^{\mu\nu}$, while the symbols $R$, $\Lambda$, and $g$ represent the Ricci-scalar, the cosmological constant, and the determinant of the metric tensor, respectively. We define the Ricci tensor as
\begin{equation} R_{\mu\nu}=\partial_\alpha\Gamma^\alpha_{\mu\nu}-\partial_\nu\Gamma^\alpha_{\mu\alpha}+\Gamma^\alpha_{\mu\nu}\Gamma^\beta_{\alpha\beta}-\Gamma^\alpha_{\mu\beta}\Gamma^\beta_{\alpha\nu}.
\end{equation}
Unless otherwise noted, our units are such that the speed of light, $c=1$; we use the metric signature $(+,-,-,-)$.

In the case of a spherically symmetric field in vacuum around a compact (point) source, we were able to derive an exact numerical solution \citep{Moffat2007e}. We found that the scalar fields $G$, $\mu$, and $\omega$ remain constant except in the immediate vicinity of the source. The spatial part of the vector field $\phi_\mu$ is zero, while its $t$-component obeys a simple exponential relationship. Meanwhile, the metric is approximately the Reissner-Nordstr\"om metric of a charged source.

Specifically, given a spherically symmetric, static metric in the standard form
\begin{equation}
d\tau^2=Bdt^2-Adr^2-r^2(d\theta^2+\sin^2\theta d\phi^2),\label{eq:schwarzschild}
\end{equation}
we found \citep{Moffat2006a,Moffat2007e} that, for a source mass\footnote{In some of our earlier work, in place of equation (\ref{eq:BB}) the formula $B\simeq 1-2G_NM/r+\omega G_0Q_5/r^2$ was used.} $M$,
\begin{align}
A&\simeq B^{-1},\\
B&\simeq 1-\frac{2G_0M}{r}+\frac{\omega G_0Q_5^2}{r^2},\label{eq:BB}\\
G&\simeq G_0=G_N+(G_\infty-G_N)\frac{M}{(\sqrt{M}+E)^2},\label{eq:G}\\
\mu&\simeq \mu_0=\frac{D}{\sqrt{M}},\label{eq:mu}\\
\omega&\simeq \omega_0=\frac{1}{\sqrt{12}},\\
\phi_t&\simeq -Q_5\frac{e^{-\mu r}}{r},
\end{align}
where $G_N$ is Newton's constant of gravitation, $Q_5=\kappa M$ is the fifth force charge associated with the source mass $M$, while $G_\infty$, $D$ and $E$ are constants. Further,
\begin{align}
\kappa&= \sqrt{\frac{G_N}{\omega}},\\
D&\simeq 6250~M_\odot^{1/2}\mathrm{kpc}^{-1},\\
E&\simeq 25000~M_\odot^{1/2},\\
G_\infty&\simeq 20G_N.
\end{align}

When $r$ is large (that is, large relative to the Schwarzschild-radius $r_S=2G_0M$ for a source mass $M$), the metric coefficients become
\begin{align}
A&\simeq B^{-1},\label{eq:A}\\
B&\simeq 1-\frac{2G_0M}{r}.\label{eq:B}
\end{align}

This is a standard result of MOG which is obtained from the MOG field equations, solved in the presence of appropriately chosen, physically motivated initial conditions \cite{Moffat2007e}.

\section{Equations of motion in a spherically symmetric field}
\label{sec:motion}

To develop an equation of motion for a point particle, and use it to derive a formula for light bending, we follow the approach presented by \cite{Weinberg1972}.

We begin with the point particle action in MOG, which is written in the form
\begin{equation}
S_\mathrm{TP}=-\int(m+\alpha\omega q_5\phi_\mu u^\mu)~d\tau=-\int(m\sqrt{g_{\alpha\beta}u^\alpha u^\beta}+\alpha\omega q_5\phi_\mu u^\mu)~d\tau,\label{eq:TPL}
\end{equation}
where $m$ is the point particle mass, $q_5$ is its fifth force charge, and $u^\alpha$ is its four-velocity. The fifth force charge is assumed to be proportional to $m$, such that $q_5=\kappa m$, where $\kappa$ is a constant. In earlier work, we determined that $\kappa=\sqrt{G_N/\omega}$. To develop an equation of motion, we compute the derivatives of the Lagrangian with respect to positions and velocities:
\begin{equation}
{\cal L}_\mathrm{TP}=-m\sqrt{g_{\alpha\beta}u^\alpha u^\beta}-\alpha\omega q_5\phi_\mu u^\mu,
\end{equation}
\begin{equation}
\frac{\partial{\cal L}_\mathrm{TP}}{\partial x^\nu}=-\frac{1}{2}mg_{\alpha\beta,\nu}u^\alpha u^\beta-\alpha\omega q_5\phi _{\mu,\nu}u^\mu-\alpha\omega_{,\nu}q_5\phi_\mu u^\mu,
\end{equation}
\begin{equation}
\frac{\partial{\cal L}_\mathrm{TP}}{\partial u^\nu}=-mg_{\alpha\nu}u^\alpha-\alpha\omega q_5\phi_\nu,
\end{equation}
\begin{equation}
\frac{d}{dt}\frac{\partial{\cal L}_\mathrm{TP}}{\partial u^\nu}=-mg_{\alpha\nu}\frac{du^\alpha}{d\tau}-mu^\beta g_{\alpha\nu,\beta}u^\alpha-\alpha\omega q_5u^\beta\phi_{\nu,\beta}-\alpha\omega_{,\beta}q_5u^\beta\phi_{\nu}.
\end{equation}
We can now construct the Euler-Lagrange equation:
\begin{equation}
\frac{\partial{\cal L}_\mathrm{TP}}{\partial x^\nu}-\frac{d}{d\tau}\frac{\partial{\cal L}}{\partial u^\nu}=mg_{\alpha\nu}\frac{du^\alpha}{d\tau}+m\Gamma_{\alpha\beta\nu}u^\alpha u^\beta+\alpha\omega q_5u^\beta(\phi_{\nu,\beta}-\phi_{\beta,\nu})-\alpha\omega_{,\nu}q_5\phi_\mu u^\mu+\alpha\omega_{,\alpha}q_5\phi_\nu u^\alpha=0,
\end{equation}
or, after rearranging terms,
\begin{equation}
m\left(\frac{du^\nu}{d\tau}+\Gamma_{\alpha\beta}^\nu u^\alpha u^\beta\right)=\alpha q_5\left[\omega u^\beta g^{\nu\alpha}B_{\alpha\beta}+\omega_{,\alpha}\phi_\mu(g^{\alpha\nu}u^\mu-g^{\mu\nu}u^\alpha)\right].
\end{equation}
This is the same as Eqs.~(31--32) of \cite{Moffat2006a}.

If $\omega$ is constant (as confirmed by the numerical solution of \cite{Moffat2007e}), we get
\begin{equation}
m\left(\frac{du^\nu}{d\tau}+\Gamma_{\alpha\beta}^\nu u^\alpha u^\beta\right)=\alpha q_5\omega u^\beta g^{\nu\alpha}B_{\alpha\beta}.
\end{equation}

The Christoffel-symbols associated with the spherically symmetric metric (\ref{eq:schwarzschild}) are

\begin{align}
\Gamma_{tt}^r=\frac{B'}{2A},&~~~~~~~~~~\Gamma_{tr}^t=\frac{B'}{2B},\\
\Gamma_{rr}^r=\frac{A'}{2A},&~~~~~~~~~~\Gamma_{r\theta}^\theta=\Gamma_{r\phi}^\phi=\frac{1}{r},\\
\Gamma_{\theta\theta}^r=-\frac{r}{A},&~~~~~~~~~~\Gamma_{\theta\phi}^\phi=\cot\theta,\\
\Gamma_{\phi\phi}^r=-\frac{r\sin^2\theta}{A},&~~~~~~~~~~\Gamma_{\phi\phi}^\theta=-\cos\theta\sin\theta.
\end{align}

The equations of motion read, using $q_5=\kappa m$ and dividing through by $m$,
\begin{align}
\frac{d^2t}{d\tau^2}+\frac{B'}{B}\frac{dt}{d\tau}\frac{dr}{d\tau}&=\alpha\kappa\omega u^\beta g^{tt}B_{t\beta},\\
\frac{d^2r}{d\tau^2}+\frac{B'}{2A}\left(\frac{dt}{d\tau}\right)^2+\frac{A'}{2A}\left(\frac{dr}{d\tau}\right)^2-\frac{r\sin^2\theta}{A}\left(\frac{d\phi}{d\tau}\right)^2&=\alpha\kappa\omega u^\beta g^{rr}B_{r\beta},\\
\frac{d^2\theta}{d\tau^2}+\frac{2}{r}\frac{dr}{d\tau}\frac{d\theta}{d\tau}-\cos\theta\sin\theta\left(\frac{d\phi}{d\tau}\right)^2&=\alpha\kappa\omega u^\beta g^{\theta\theta}B_{\theta\beta},\\
\frac{d^2\phi}{d\tau^2}+\frac{2}{r}\frac{dr}{d\tau}\frac{d\phi}{d\tau}+2\cot\theta\frac{d\theta}{d\tau}\frac{d\phi}{d\tau}&=\alpha\kappa\omega u^\beta g^{\phi\phi}B_{\phi\beta}.
\end{align}
We can set $\theta=\pi/2$ without loss of generality. We can also recognize that the only non-zero components of $B_{\alpha\beta}$ are $B_{tr}=-B_{rt}=\partial_t\phi_r-\partial_r\phi_t=-\phi_t'$. We get
\begin{align}
\frac{d^2t}{d\tau^2}+\frac{B'}{B}\frac{dt}{d\tau}\frac{dr}{d\tau}&=-\frac{\alpha\kappa\omega}{B}\frac{dr}{d\tau}\phi_t',\label{eq:t}\\
\frac{d^2r}{d\tau^2}+\frac{B'}{2A}\left(\frac{dt}{d\tau}\right)^2+\frac{A'}{2A}\left(\frac{dr}{d\tau}\right)^2-\frac{r}{A}\left(\frac{d\phi}{d\tau}\right)^2&=-\frac{\alpha\kappa\omega}{A}\frac{dt}{d\tau}\phi_t',\label{eq:r}\\
\frac{d^2\phi}{d\tau^2}+\frac{2}{r}\frac{dr}{d\tau}\frac{d\phi}{d\tau}&=0.\label{eq:phi}
\end{align}
Eq.~(\ref{eq:t}) can be rearranged after multiplying with $B$:
\begin{equation}
B\frac{d^2t}{d\tau^2}+\frac{dB}{d\tau}\frac{dt}{d\tau}=-\alpha\kappa\omega\frac{d\phi_t}{d\tau},
\end{equation}
which can be integrated to yield
\begin{equation}
\frac{dt}{d\tau}=\frac{C-\alpha\kappa\omega\phi_t}{B},\label{eq:dtdt}
\end{equation}
where $C$ is a constant. Since $B\rightarrow 1$ and $\phi_t\rightarrow 0$ as $r\rightarrow\infty$, an asymptotically flat spacetime requires $C=1$.

Eq.~(\ref{eq:phi}) can be integrated directly:
\begin{equation}
r^2\frac{d\phi}{d\tau}=J,\label{eq:dphidtau}
\end{equation}
where $J$ is a constant of integration, which we identify as the angular momentum per unit mass.

Using these results in Eq.~(\ref{eq:r}) yields
\begin{equation}
\frac{d^2r}{d\tau^2}+\frac{A'}{2A}\left(\frac{dr}{d\tau}\right)^2-\frac{J^2}{Ar^3}+\frac{B'}{2A}\left(\frac{1-\alpha\kappa\omega\phi_t}{B}\right)^2=-\frac{\alpha\kappa\omega(1-\alpha\kappa\omega\phi_t)}{AB}\phi_t'.
\end{equation}
In the case of general relativity, $\alpha$ and $\phi_t$ are zero, and the particle moves along a geodesic. This is not the case here, but we can still integrate our equation. Multiplication by $2Adr/d\tau$ leads to
\begin{equation}
\frac{d}{d\tau}\left[A\left(\frac{dr}{d\tau}\right)^2+\frac{J^2}{r^2}-\frac{(1-\alpha\kappa\omega\phi_t)^2}{B}\right]=0.
\end{equation}
Integration yields
\begin{equation}
A\left(\frac{dr}{d\tau}\right)^2+\frac{J^2}{r^2}-\frac{(1-\alpha\kappa\omega\phi_t)^2}{B}=-{\cal E},\label{eq:E}
\end{equation}
where ${\cal E}$ is another constant of integration. After using (\ref{eq:dtdt}), we get
\begin{equation}
\frac{A}{B^2}\left(\frac{dr}{dt}\right)^2+\frac{J^2}{(1-\alpha\kappa\omega\phi_t)^2}\frac{1}{r^2}-\frac{1}{B}=-\frac{{\cal E}}{(1-\alpha\kappa\omega\phi_t)^2}.
\end{equation}
This is an exact result. From this result, we can develop the equation of motion for a non-relativistic particle in the usual form. For this, let us assume that we are far from a source, and the metric is that of Schwarzschild, in accordance with (\ref{eq:A}) and (\ref{eq:B}). Then, our equation of motion becomes
\begin{equation}
\frac{1}{(1-2GM/r)^3}\left(\frac{dr}{dt}\right)^2+\frac{J^2}{(1-\alpha\kappa\omega\phi_t)^2}\frac{1}{r^2}-\frac{1}{1-2GM/r}=\frac{-{\cal E}}{(1-\alpha\kappa\omega\phi_t)^2}.
\end{equation}
Multiplying both sides with $(1-2GM/r)^3$ gives
\begin{equation}
\left(\frac{dr}{dt}\right)^2+\frac{J^2}{(1-\alpha\kappa\omega\phi_t)r^2}\left(1-\frac{2GM}{r}\right)^3-\left(1-\frac{2GM}{r}\right)^2=\frac{-{\cal E}}{\left(1-\alpha\kappa\omega\phi_t\right)^2}\left(1-\frac{2GM}{r}\right)^3.
\end{equation}
Fully differentiating with respect to $t$, dividing through with $2dr/dt$ and then rearranging terms yields
\begin{equation}
\frac{d^2r}{dt^2}-\frac{(1-2GM/r)^3}{1-\alpha\kappa\omega\phi_t}\frac{J^2}{r^3}+\frac{(1-2GM/r)^2}{2(1-\alpha\kappa\omega\phi_t)^2}\frac{J^2}{r^2}\alpha\kappa\omega\phi_t'+\frac{(1-2GM/r)^2}{1-\alpha\kappa\omega\phi_t}\frac{3J^2GM}{r^4}-\left(1-\frac{2GM}{r}\right)\frac{2GM}{r^2}
$$ $$
=\frac{-{\cal E}(1-2GM/r)^3}{(1-\alpha\kappa\omega\phi_t)^3}\alpha\kappa\omega\phi_t'-\frac{3{\cal E}(1-2GM/r)^2}{(1-\alpha\kappa\omega\phi_t)^2}\frac{GM}{r^2}.\label{eq:47}
\end{equation}
From (\ref{eq:E}), taking the large-$r$ limit, we get
\begin{equation}
{\cal E}=1-v^2,
\end{equation}
where $v=dr/d\tau\simeq dr/dt$ is the velocity of the particle at infinity. For photons, ${\cal E}=0$, for material particles in unbound orbits, $0<{\cal E}<1$, and for bound particles, ${\cal E}\ge 1$.

For a non-relativistic particle, $v^2\ll 1$. Further, in the weak field limit, $1-2GM/r\simeq 1$, $1-\alpha\kappa\omega\phi_t\simeq 1$, and $J^2/2r^2\ll 1$, so (\ref{eq:47}) becomes
\begin{equation}
\frac{d^2r}{dt^2}-\frac{J^2}{r^3}=-\alpha\kappa\omega\phi_t'-\frac{GM}{r^2}.
\label{eq:nonrel}
\end{equation}
This equation of motion is a direct consequence of the solution of the field equations presented in Section~\ref{sec:theory}, used in combination with the postulated test particle action (\ref{eq:TPL}).

\section{The bending of light}
\label{sec:bending}

To calculate light bending, we return to the exact result in (\ref{eq:E}). The first step is eliminating $d\tau$ using (\ref{eq:dphidtau}), since we are interested in the shape of the orbit, not its time evolution:
\begin{equation}
A\left(\frac{dr}{d\phi}\frac{d\phi}{d\tau}\right)^2+\frac{J^2}{r^2}-\frac{(1-\alpha\kappa\omega\phi_t)^2}{B}=-{\cal E},
\end{equation}
or,
\begin{equation}
\frac{A}{r^4}\left(\frac{dr}{d\phi}\right)^2+\frac{1}{r^2}=-\frac{{\cal E}}{J^2}+\frac{(1-\alpha\kappa\omega\phi_t)^2}{J^2B}.\label{eq:drdphi}
\end{equation}
From this we get
\begin{equation}
\phi=\pm\int\frac{A^{1/2}}{r^2\left(\frac{(1-\alpha\kappa\omega\phi_t)^2}{J^2B}-\frac{{\cal E}}{J^2}-\frac{1}{r^2}\right)^{1/2}}~dr.\label{eq:phir}
\end{equation}

At closest approach to a source, $r=r_0$ and $dr/d\phi$ vanishes. Then, (\ref{eq:drdphi}) becomes
\begin{equation}
J=r_0\sqrt{\frac{[1-\alpha\kappa\omega\phi_t(r_0)]^2}{B}+v^2-1}.
\end{equation}
Putting this into (\ref{eq:phir}) gives
\begin{equation}
\phi=\phi_\infty+\int_r^\infty{\frac{A^{1/2}}{r^2\left[\frac{1}{r_0^2}\left\{\frac{[1-\alpha\kappa\omega\phi_t]^2}{B}+v^2-1\right\}\left\{\frac{[1-\alpha\kappa\omega\phi_t(r_0)]^2}{B(r_0)}+v^2-1\right\}^{-1}-\frac{1}{r^2}\right]^{1/2}}}~dr.
\end{equation}
The deflection for a particle coming from infinity to $r_0$ and then off to infinity is twice this angle:
\begin{equation}
\Delta\phi=2|\phi-\phi_\infty|-\pi.
\end{equation}
For a photon, $v=1$ and
\begin{equation}
\Delta\phi_\gamma=2\left|\int_{r_0}^\infty{\frac{A^{1/2}}{r^2\left[\frac{1}{r_0^2}\left\{\frac{[1-\alpha\kappa\omega\phi_t]^2}{B}\right\}\left\{\frac{[1-\alpha\kappa\omega\phi_t(r_0)]^2}{B(r_0)}\right\}^{-1}-\frac{1}{r^2}\right]^{1/2}}}~dr\right|-\pi.
\end{equation}
In the case of weak fields, $1-\alpha\kappa\omega\phi_t\simeq 1$ and we get
\begin{equation}
\Delta\phi_\gamma=2\left|\int_{r_0}^\infty{\frac{1}{r}\left[\frac{r^2}{r_0^2}\frac{B(r_0)}{AB}-\frac{1}{A}\right]^{-1/2}}~dr\right|-\pi.
\end{equation}
This formula is formally identical to the light bending formula in the weak field limit of general relativity, with one notable difference: instead of $G=G_N$, we are using $G=G_\infty=(1+\alpha)G_N$ in the Schwarzschild coefficients $A$ and $B$. From this formula, the approximate deflection can be calculated as \citep{Moffat2006a,Weinberg1972}:
\begin{equation}
\Delta\phi_\gamma=\frac{4GM}{r_0}=\frac{4(1+\alpha)G_NM}{r_0}.\label{eq:bend}
\end{equation}

If $\phi_t$ cannot be ignored, we can use the form \citep{Moffat2007e}:
\begin{equation}
\phi_t=-Q_5\frac{e^{-\mu r}}{r}=-\kappa M\frac{e^{-\mu r}}{r},
\end{equation}
where $Q_5=\kappa M$ is the fifth force charge of the source with mass $M$. This yields the formula for light bending in the strong field of a point source in the form
\begin{equation}
\Delta\phi_\gamma=2\left|\int_{r_0}^\infty{\frac{1}{r}\left[\frac{[r+\alpha G_NM\exp(-\mu r)]^2}{[r_0+\alpha G_NM\exp(-\mu r_0)]^2}\frac{B(r_0)}{AB}-\frac{1}{A}\right]^{-1/2}}~dr\right|-\pi,
\end{equation}
where we used $\kappa^2\omega=G_N$.

\twocolumn
\section{Extended source distributions}
\label{sec:extended}

The nonrelativistic equation of motion (\ref{eq:nonrel}) can be further simplified when only radial motion is considered, such that $J=0$:
\begin{equation}
\ddot{r}=-\frac{G_NM}{r^2}\left[1+\alpha-\alpha(1+\mu r)e^{-\mu r}\right].
\end{equation}
This corresponds to the potential
\begin{equation}
\Phi(r)=-\frac{G_NM}{r}\left[1+\alpha-\alpha e^{-\mu r}\right],
\end{equation}
such that ($r=|\vec{r}|$):
\begin{equation}
\ddot{\vec{r}}=-\nabla\Phi(\vec{r}).
\end{equation}
We write the potential as the sum of two constituents:
\begin{equation}
\Phi(\vec{r})=\Phi_N(\vec{r})+\Phi_Y(\vec{r}),
\end{equation}
where
\begin{equation}
\Phi_N(\vec{r})=-\frac{(1+\alpha)G_N M}{r}
\end{equation}
and
\begin{equation}
\Phi_Y(\vec{r})=\frac{\alpha G_NM}{r}e^{-\mu r}.\label{eq:66}
\end{equation}
We find that $\Phi_N$ is the solution of the Poisson equation \citep{Brownstein2007,Moffat2007c}:
\begin{equation}
\nabla^2\Phi_N(\vec{r})=4\pi(1+\alpha)G_N\rho(\vec{r}),\label{eq:PhiN}
\end{equation}
with $\rho(\vec{r})=M\delta^3(\vec{r})$, where $\delta$ is Dirac's delta function.

For $\Phi_Y$, we consider the inhomogeneous Helmholtz equation \citep{Brownstein2007,Moffat2007c} in the unknown function $f(\vec{r})$:
\begin{equation}
(\nabla^2+k^2)f(\vec{r})=-\delta^3(\vec{r}).
\end{equation}
This equation is solved by
\begin{equation}
f(\vec{r})=\frac{e^{ikr}}{4\pi r}.
\end{equation}
Replacing $f$ with $\Phi_Y$ and $k$ with $i\mu$, and using (\ref{eq:66}) we find that
\begin{equation}
(\nabla^2-\mu^2)\Phi_Y(\vec{r})=-4\pi\alpha G_N\rho(\vec{r}).\label{eq:PhiY}
\end{equation}
Adding (\ref{eq:PhiN}) and (\ref{eq:PhiY}), we get
\begin{equation}
\nabla^2\Phi(\vec{r})=4\pi G_N\rho(\vec{r})+\mu^2\Phi_Y(\vec{r}).\label{eq:MOGP}
\end{equation}
If $\mu=0$, we get back the Poisson equation for Newtonian gravity, as expected.

If $\rho$ is not a point source but a general continuous distribution of matter, the solution of the inhomogeneous Helmholtz equation is in the form  \citep{Brownstein2007,Moffat2007c}:
\begin{equation}
\Phi_Y(\vec{r})=\alpha G_N\int\frac{e^{-\mu|\vec{r}-\tilde{\vec{r}}|}}{|\vec{r}-\tilde{\vec{r}}|}\rho(\tilde{\vec{r}})~d^3\tilde{\vec{r}}.
\end{equation}
Thus,
\begin{equation}
\nabla^2\Phi(\vec{r})=4\pi G_N\rho(\vec{r})+\alpha\mu^2G_N\int\frac{e^{-\mu|\vec{r}-\tilde{\vec{r}}|}}{|\vec{r}-\tilde{\vec{r}}|}\rho(\tilde{\vec{r}})~d^3\tilde{\vec{r}}.\label{eq:nabla2}
\end{equation}
Or, since $\ddot{\vec{r}}=-\nabla\Phi(\vec{r})$, we can write
\begin{equation}
\nabla\cdot\ddot{\vec{r}}=-4\pi G_N\rho(\vec{r})-\alpha\mu^2G_N\int\frac{e^{-\mu|\vec{r}-\tilde{\vec{r}}|}}{|\vec{r}-\tilde{\vec{r}}|}\rho(\tilde{\vec{r}})~d^3\tilde{\vec{r}}.
\end{equation}

In the case of a point source with mass $M$, $\mu$ is given by (\ref{eq:mu}), while $\alpha=(G_0-G_N)/G_N$ is determined by (\ref{eq:G}):
\begin{equation}
\alpha=\frac{M}{(\sqrt{M}+E)^2}\left(\frac{G_\infty}{G_N}-1\right).\label{eq:alpha}
\end{equation}

\begin{figure}
\centering\includegraphics[width=0.5\linewidth]{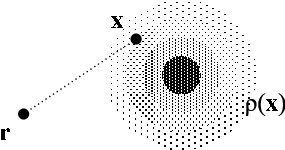}
\caption{The position $\vec{r}$ of a test particle relative to a point $\vec{x}$ inside an extended mass distribution characterized by $\rho(\vec{x})$.}
\label{fig:rho}
\end{figure}

\begin{figure*}
\centering\includegraphics[width=0.75\linewidth]{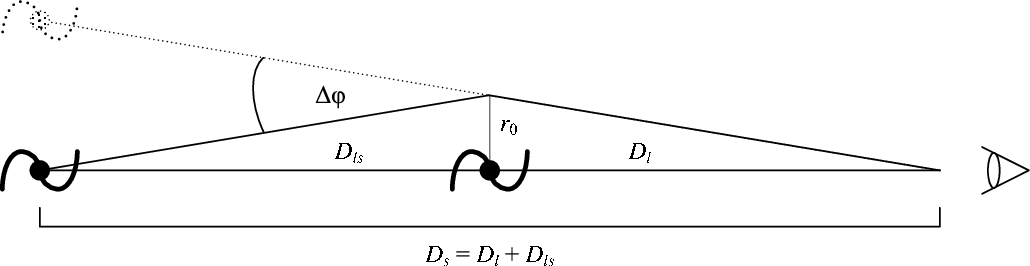}
\caption{The geometry of thin lensing: a foreground object, at distance $D_l$ from the observer, deflects light from a background object, at distance $D_s$, by an angle $\Delta\phi$.}
\label{fig:lensing}
\end{figure*}

Clearly, in the case of an extended mass distribution, these expressions must be modified. Presently, we do not have solutions to the MOG field equations for extended matter distributions. Therefore, we treat this problem phenomenologically. We seek an effective mass function ${\cal M}(\vec{x},\vec{r})$, to be used in place of $M$ in (\ref{eq:alpha}) and (\ref{eq:mu}), that determines an ``effective mass'', to be used in the formulae for $\alpha$ and $\mu$, when calculating the gravitational influence of matter at $\vec{x}$ on a test particle located at $\vec{r}$ (see Figure~\ref{fig:rho}). The function should yield the mass of the source in the case of a point source, and a mass proportional to volume in the case of a constant distribution. One function that satisfies these criteria is in the form,
\begin{equation}
{\cal M}(\vec{x},\vec{r})=\int\rho(\bar{\vec{x}})\exp\left(-\xi\frac{|\bar{\vec{x}}-\vec{x}|}{|\vec{r}-\vec{x}|}\right)~d^3\bar{\vec{x}},\label{eq:76}
\end{equation}
with the coefficient $\xi$ to be determined from observation, e.g., by comparison with the Bullet Cluster data. For a constant distribution $\rho(\vec{x})=\rho_0$, this function gives ${\cal M}(\vec{x},\vec{r})\propto |\vec{r}-\vec{x}|^3$.
If $\rho$ is not constant, we get the following expressions for $\alpha$ and $\mu$:
\begin{align}
\alpha(\vec{x},\vec{r})&=\frac{{\cal M}(\vec{x},\vec{r})}{(\sqrt{{\cal M}(\vec{x},\vec{r})}+E)^2}\left(\frac{G_\infty}{G_N}-1\right),\\
\mu(\vec{x},\vec{r})&=\frac{D}{\sqrt{{\cal M}(\vec{x},\vec{r})}}.
\end{align}
This means that in (\ref{eq:nabla2}), we must now move $\alpha$ and $\mu$ under the integral sign: \begin{equation}
\nabla^2\Phi(\vec{r})=4\pi G_N\rho(\vec{r})+G_N\int\alpha(\tilde{\vec{r}},\vec{r})\mu^2(\tilde{\vec{r}},\vec{r})\frac{e^{-\mu(\tilde{\vec{r}},\vec{r})|\vec{r}-\tilde{\vec{r}}|}}{|\vec{r}-\tilde{\vec{r}}|}\rho(\tilde{\vec{r}})~d^3\tilde{\vec{r}}.
\end{equation}

Given that ${\cal M}$ depends only on $\vec{x}$ and $r_0=|\vec{x}-\vec{r}|$, we can also write (\ref{eq:76}) in the form
\begin{equation}
{\cal M}(\vec{x},r_0)=\int\rho(\bar{\vec{x}})\exp\left(-\xi\frac{|\bar{\vec{x}}-\vec{x}|}{r_0}\right)~d^3\bar{\vec{x}},
\end{equation}
and $\alpha$ and $\mu$ can also be written as functions of $\vec{x}$ and $r_0$.

For lensing, we can use the $\kappa$-convergence formula \citep{Peacock1999,Weinberg2008}:
\begin{equation}
\kappa(x,y)=\int\frac{4\pi G(\vec{x})}{c^2}\frac{D_lD_{ls}}{D_s}\rho(\vec{x})~dz,\label{eq:kappa}
\end{equation}
where $D_l$ is the distance from the observer to the lensing plane, $D_s$ is the distance from the observer to the distant light source, $D_{ls}=D_s-D_l$ (see Figure~\ref{fig:lensing}), and a rectilinear coordinate system is used such that $\vec{x}=(x,y,z)$, and the $z$-axis coincides with the line connecting the observer with the distant light source. In our case, $G$ is given by
\begin{equation}
G(\vec{x})=[1+\alpha(\vec{x},r_0)]G_N,
\end{equation}
where $r_0$ now represents the distance between the point where a straight line connecting the light source with the observer intersects the lensing plane, and the point where the bent ray of light intersects the lensing plane. For a point mass, this distance can be calculated as
\begin{equation}
r_0=\frac{D_lD_{ls}}{D_s}\sin{\Delta\phi}\simeq\frac{D_lD_{ls}}{D_s}\Delta\phi.\label{eq:b}
\end{equation}
In the case of extended distributions, $\Delta\phi$ can be calculated by
\begin{equation}
\Delta\phi(x,y)=\frac{2}{c^2}\left|\int \nabla_\perp\Phi(\vec{x})~dz\right|,\label{eq:bb}
\end{equation}
where $\nabla_\perp$ represents the gradient operator in the $(x,y)$ lensing plane. $\Phi(\vec{x})$ would, in the case of general relativity, be given by the Poisson equation $\nabla^2\Phi(\vec{x})=4\pi G_N\rho(\vec{x})$. Our result for light bending (\ref{eq:bend}) makes it clear that in the case of MOG, we cannot use the nonrelativistic equation (\ref{eq:MOGP}). Instead, the correct equation to use for the effective potential $\Phi$ is
\begin{equation}
\nabla^2\Phi(\vec{x})=4\pi [1+\alpha(\vec{x},r_0)]G_N\rho(\vec{x}),
\end{equation}
or, after substituting (\ref{eq:b}) and (\ref{eq:bb}), we get
\begin{equation}
\nabla^2\Phi(\vec{x})=4\pi G_N\rho(\vec{x})\left[1+\alpha\left(\vec{x},\frac{D_lD_{ls}}{D_s}\frac{2}{c^2}\left|\int{\nabla_\perp\Phi(\vec{x})}~dz\right|\right)\right],
\label{eq:PhiExt}
\end{equation}
an equation that is solvable for $\Phi$.

This equation (\ref{eq:PhiExt}) was derived using the MOG field equations presented in Section~\ref{sec:theory}, using the postulated point particle Lagrangian (\ref{eq:TPL}) and a phenomenological effective mass function (\ref{eq:76}) used to represent extended mass distributions.

\section{Conclusions}
\label{sec:conclusions}

We derived an exact formulation for light bending in the spherically symmetric field of a point source in modified gravity (MOG). Introducing a phenomenological approach to model extended distributions of matter, we showed how the theory can be used to compute lensing. This work is directly applicable to the case of the Bullet Cluster 1E0657-558. In previous work \citep{Brownstein2007}, this cluster was studied using the nonrelativistic equations of motion in MOG. These equations predict a different deflection angle near sources, as unlike light rays, nonrelativistic particles are influenced by the MOG fifth force and do not travel along geodesics. Further out from sources ($r\gg \mu^{-1}$), the two predictions agree. This condition is satisfied for most of the light rays that are deflected by the gravity of the Bullet Cluster, validating our previous result.

The merging of three clusters of galaxies in Abell 520 has created a mystery regarding the nature of dark matter and its role in merging clusters \citep{Mahdavi2007}. In contrast to the Bullet Cluster, the data show that the dark matter does not separate from the baryon plasma situated at approximately the center of the merging clusters, whereas the galaxies and stars have separated to the sides of the cluster. Moreover, these galaxies do not reveal the existence of dark matter haloes. In our derivation of the $\kappa$-convergence lensing, we believe that MOG can provide a natural explanation for these results without exotic dark matter, by modifying the mass profile at the center of the merging clusters. This will be the subject of future research.

The phenomenological formalism we developed for modeling extended distributions of matter may also be useful in future work, including $N$-body simulations using MOG, and cosmological computations.

\section*{Acknowledgements}

We thank Joel Brownstein for helpful discussions. The research was partially supported by National Research Council of Canada. Research at the Perimeter Institute for Theoretical Physics is supported by the Government of Canada through NSERC and by the Province of Ontario through the Ministry of Research and Innovation (MRI).

\bibliography{refs}
\bibliographystyle{mn2e}

\end{document}